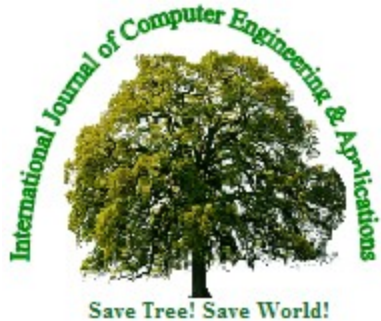

# A NOVEL APPROACH FOR HYBRID DATABASES

Rajesh Kumar Tiwari[1]

[1]R.V.S. College of Engineering and Technology, Jamshedpur, Jharkhand, India

**ABSTRACT**

In the current world of economic crises, the cost control is one of the chief concerns for all types of industries, especially for the small venders. The small vendors are suppose to minimize their budget on Information Technology by reducing the initial investment in hardware and costly database servers like ORACLE, SQL Server, SYBASE, etc. for the purpose of data processing and storing. In other divisions, the electronic devices manufacturing companies want to increase the demand and reduce the manufacturing cost by introducing the low cost technologies. The new small devices like ipods, iphones, palm top etc. are now-a-days used as data computation and storing tools.

For both the cases mentioned above, instead of going for the costly database servers which additionally requires extra hardware as well as the extra expenses in training and handling, the flat file may be considered as a candidate due to its easy handling nature, fast accessing, and of course free of cost. But the main hurdle is the security aspects which are not up to the optimum level. In this paper, we propose a methodology that combines all the merit of the flat file and with the help of a novel steganographic technique we can maintain the utmost security fence. The new proposed methodology will undoubtedly be highly beneficial for small vendors as well as for the above said electronic devices manufacturer

**Keyword-** Relational Database, Attribute, Cryptography, Hybrid, Security, Stego-Key, Flat file.

## 1. INTRODUCTION

Computers are virtually now becoming a part of every activity in business. The most common and obvious applications are business applications, such as, keeping records of transactions, airline reservation status, purchasing good through e-commerce, or the amount of goods that are available in the stores. They are also used in design problems, such as designing a building or setting a project schedule. In addition, computers are now accessibly used in the data processing and storing.

Digital data are widely distributed and extensible used by the data processing units. Since digital data can easily be duplicated and modified, there is great deal of concern about the integrity and protection of data.

As a part of solution to the above kind of problems, the concept of steganography has considerably been given much attention in recent years. The word steganography is a Greek word giving a meaning to it as 'writing in hiding'. The main purpose of steganography is to hide data in a cover media so that other will not notice [1, 2, 3]. The characteristics of cover media depends on the amount of data that can be hidden, the perceptibility of the message and its robustness [4, 5, 6, 8, 10, 11]. New digital technology gives new way to apply steganographic techniques including hiding information in image, audio, video, html, text and executables files. [7,9,13,14, 15].

## 2. RELATED WORK

As we know, a flat file is a plain text or mixed text binary files which usually contains one record per line. Each record of the flat file database can be separated by delimiters. The different delimiters may be a space, commas, tab, semicolons etc. There are no structural relationships between the records (Fig.1).

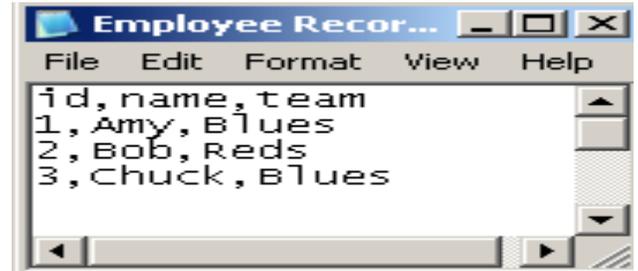

Fig.1 Flat file

The data storing and processing may be done either directly or by using the application program (Fig.2).

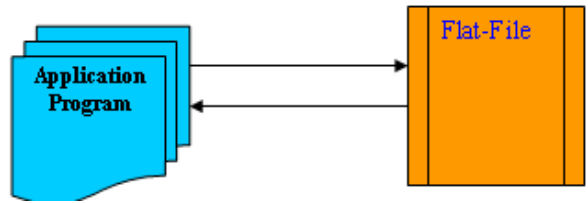

Fig.2 Basic model of flat files database application

One of the major drawbacks of the flat file database is the less security level, without the interference of the application program one can easily operate it.

In the relational database system, the database driver plays a key role between the application program and the database server as shown in fig.3. This makes the process slow as compare to that of flat file system. Further, the high cost, training expenses and other database admiration requirements are considered as some of the hurdles among others.

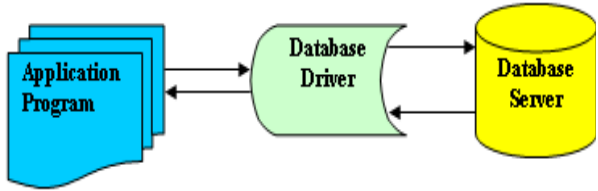

Fig.3 Basic model of relational database application

In the other side, the most suitable cover media for Steganography is an image and numerous methods have been discussed in this aspect. The main reason behind it is the large redundant and the possibility of hiding information in the image without attracting attention to the human visual system. In this respect, some techniques have already been developed by the authors among others [1, 2, 3] using features like

i. Substitution
ii. Masking & Filtering
iii. Transformation Technique

In the process of data hiding, the method of substitution generally does not increase the size of the file, depending on the size of the hidden message it can eventually cause a noticeable change from the unmodified version of the respective file [1, 2, 4, 6]. In this regard, the Least Significant Bit (LSB) insertion technique has been considered as an approach for embedding information in an image file. In this case, every least significant bit of some or all of the bytes inside an image is changed to a bit of the secret message.

In case of context masking and filtering techniques, we first start analyzing the image then find the significant areas where the hidden message will be used to cover the image. Finally we embed the data in that particular area. Therefore, while in the case of LSB techniques all the least significant bits are changed, in the case of masking and filtering techniques we just say that the changes can take place only in selected areas.

In addition to the above two approaches for message hiding, the third transform techniques has also some significance in embedding the message by modulating coefficients in a transform domain. As an example, Discrete Cosine Transform works by using quantization on the least important parts of the Image in respect to the human visual capabilities. Marvel [6] has proposed a spread spectrum image steganographic technique for image carrier and able to conceal 5 kB of secret data in 512 x 512 image size. Julio C. Hernandez-Castro [7] has given a concept for steganography in games where the less robust technique saves only few parts of total carrier files. Raja [8] concluded that three secret bits can be embedded in one pixel. Amin [10] has given a secured information hiding system which can embed 60 kB message in 800 x 600 pixels image. We can say here that all of the above methods used for embedding secret data in an image file works on the principle of replacing the entire or some chosen pixels. Due to this replacement policy, however, we are not able to embed the voluminous data.

The propose model inherits all the basic properties of the flat file and with the help of new high capacity steganographic techniques. The new hybrid model maintains the high level of security also.

## 3. PROPOSED APPROACH

In the propose hybrid database model, we are highlighting in mainly two areas, enhancement of the storing capacity for the steaganographic carrier and increasing the data storing, accessing and security. Like the plain text flat file, the image file also consists of the binary data so; here we are taking image file for storing the data. For simplicity, we have divided the process in two steps. The first process takes care for steganographic concept for storing the voluminous data into the image file. Where as, the second process fulfill other requirement of the hybrid database system.

**Step-I: Using steganographic concept**

Steganography is a science of hiding the data into the cover file. A digital image file is being used as a cover file for the storage purpose of the database. Digital images are commonly of two types 8 bit images and 24 bit images. These images can be further logically divided into two parts namely primary part and secondary part. While the primary part consists of all preliminary information like file name, file type, compression type etc., the secondary part keeps the required information of the pixels being used. So, without disturbing the primary and secondary part, we can append the data at the end of the secondary part of the image file. This leads to get following advantages: first there will be hardly any visual changes in the image file and secondly, a small image file may be sufficient to keep a large amount of data fig. 4.

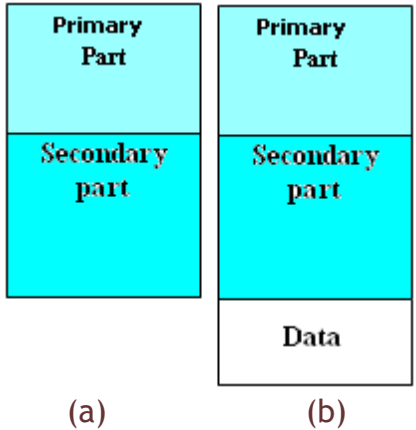

(a)　　　　　(b)

Fig. 4. (a) Image Structure (b) Image Structure with data.

For the above purpose, the required pseudo code is given below.

/* Pseudo Code for Data Storing */

Data_Storing
{
1DArray Target_Image= Load ("Cover Frame", &length);
1D Array Stego_Image [ ];
1D Array Data [ ];
Integer I, J=0, k;
Input Data;

For I =1 to Length (Target_Image)
    Stego_Image[i] = Target_Image[i];
End For

For I =1 to Length (Data [ ])
    Stego_Image[i] = Data[i];
End For

SaveImage ("Stego_Image");
Rename_Image ("Target_Image", "Stego_Image");
Remove_Image ( "Stego_Image");
}

Step-II: Modeling into a hybrid database

The steganography is a tool which can be used for high security fence. In the proposed hybrid database system we have combined the basic feature of flat file and relational database system.

The relational database system uses the terms table, row and column for the relational model terms relation, tuple, and attributes. The data arrangement consist series of columns and rows organized into a tabular format and each rows and columns of the table meets the standard definition. In our hybrid database system we are marinating the same concept by taking the each relation in one image file (i.e. one image file for one table). We are storing relation name (table name), attributes (columns) and their category (data types) in the secondary part of the target image by using the least significant techniques. Each attributes in image file is

International Journal of Computer Engineering & Applications

ordinarily restricted to a specific primary data types like character, float, money, Boolean, etc. Just for illustration purpose we are taking one example. The relation *R* having *n* attributes and a primary key attributes *P* can be denoted as, *R (P, $A_1$, $A_2$ ....$A_n$)*. In the above more than one primary key attribute can also be considered for the required purpose and the relation R can be stored in the secondary part of the target image file fig. 5. The pseudo code for storing the table structure in the target image file is given below.

/* Pseudo Code for Table Structure Storing*/
Table_Structure_Storing
{
1DArray Target_Image= Load ("Cover Frame", &length);
1D Array Stego_Image [ ];
1D Array Table_Structure [ ];
1D Array ASCII_data [ ];
Integer I, J=0, k;
Input Table_Structure;
/*Secret Data bit Conversion by using predefined or library functions */
    Bit_Conversion (ASCII_data [ ], Table_Structure [ ]);
For I =1 to Length (ASCII_data [ ])
    If  ASCII_data [I]  <>  MOD (Target_Image[k], 2) then
      If ASCII_data [I] =0 Then
            Target_Image[k] = Asc (Target_Image[k])-1;
      Else
            Target_Image[k] = Asc (Target_Image[k]) +1;
      End If;
 End If;
 k=k+1;
End For
SaveImage ("Stego_Image");
}

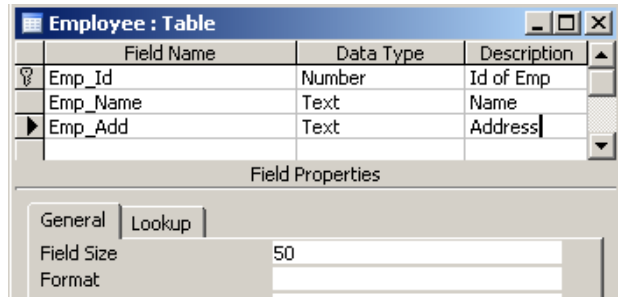

(a)

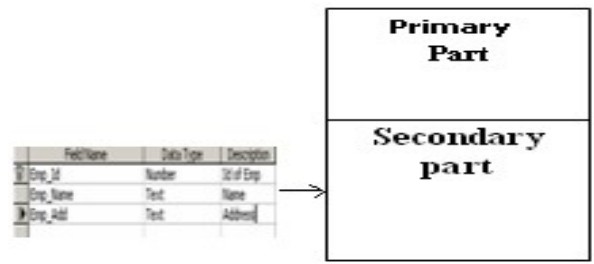

(b)

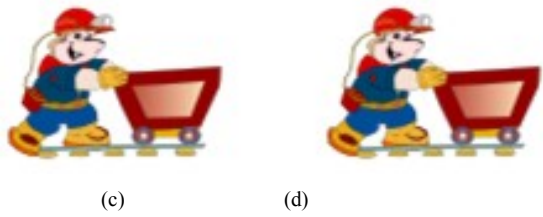

(c)          (d)

Fig. 5. (a) Table Structure (b) Image with table structure (c ) Target Image. (d) Image with Table structure.

Database integrity refers to the requirement that information be protected from improper modification. Modification of data includes creation of new columns or rows, insertion of new record, modification in any or some of pre-existing records or the deletion of the old records. Integrity is lost if unauthorized changes are made to the data intentionally or by accidental acts. If the loss of system or data integrity is not corrected in due course of time, further continuous use of the impure system or corrupted data could result in inaccuracy, fraud or erroneous decision. Therefore, by storing the basic table structure with the attributes constraint in the secondary part of the image file, we are maintaining the utmost level of security in our proposed hybrid database system.

The referential integrity is one of the prime features of the relational database system; it is specified via the FOREIGN KEY clause, as shown in fig. 6.

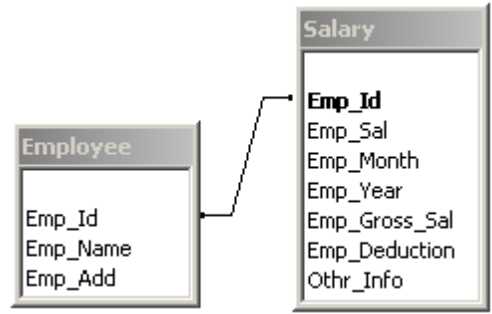

Fig. 6. Parent Child relation at Table level.

However, a referential integrity constraint can be violated when tuples are inserted or deleted or when a foreign key attribute value is modified. In the proposed hybrid database system, we have established the foreign key concept by the help of an extra image file. Since we store each relation in a separate file with the attribute level constraint, the parent-child relation will require three image files, one for the parent table, one for the child table and the last for storing the constraint table level. For example, if we have two tables say, employee master and employee salary. The employee master table contains the primary information of the employee and the employee salary table holds the salary information of the corresponding employee of the employee master table. There is parent child relation between the two tables. We store the foreign key constraint in the secondary part of third image file as shown in fig.7.

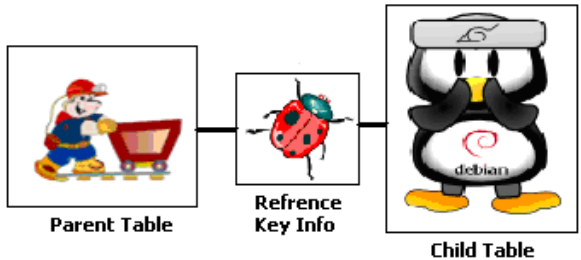

**Fig. 7. (a) Parent Child relation for Hybrid database model.**

## 4. IMPLEMENTATION

Based on the above methodologies we have designed our hybrid database system in Microsoft platform, and it may however be implemented in other programming platforms like JAVA & C. The proposed system gives the option for selecting the image file. Here it is important to mention that our proposed model stores data in all types of image files irrespective of their sizes. The next step is for creating the attributes and the corresponding basic data types with the attribute label constraint. All information of the basic structure of the table are being stored at the secondary part of the target image as shown in the following fig. 8.

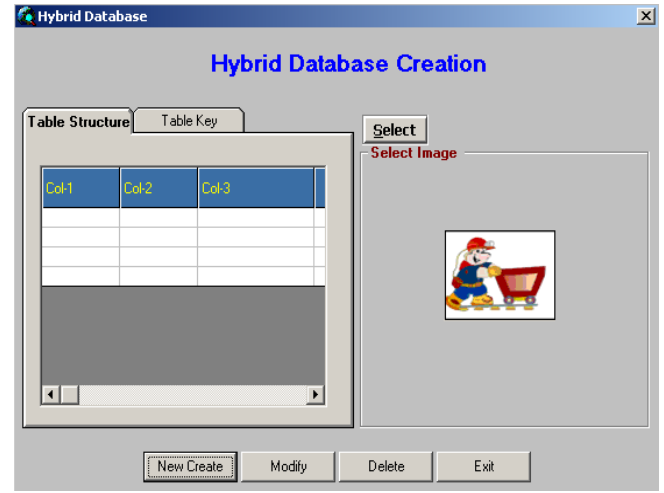

Fig. 8. (a) Table Creation.

The referential constraint can be maintained by selecting three different types of image files each for parent, child and their relations.

The second program takes care of the creation of new records, modification and deletion of pre-existing records. It takes the input from the user for the different types of operations. For the creation of the new record, it takes the structure from the secondary part of the image file and displays the entire fields with their data type. Based on the data type and the constraints we insert the record in the image file and similarly, for the modification and deletion of records, we check all the constraint including the foreign key constraint.

## 5. Conclusion

The suitability of steganography as a tool to conceal highly sensitive information into image file has been discussed. By using this new proposed methodology we can remove the major drawbacks of the flat file system. The proposed method lies between the flat file system and the expensive relational database system. It is important to mention here that we can use this model in place of the flat file system as well as in the electronic devices like ipods, iphones, palm top etc. as these devices now a-days are being used as a data editing and storing tool.

<p><s>International Journal of Computer Engineering & Applications</s></p>